\documentclass[aps,prb,twocolumn,showpacs,a4paper,superscriptaddress]{revtex4-1}
\usepackage{graphicx,bm,amsmath,dcolumn,amssymb}
\usepackage{longtable}
\graphicspath{{figures}}
\usepackage{geometry,hyperref,color}
\begin{document}
\title{Emergent O($n$) Symmetry in a Series of 3D Potts Models}
  \author{Chengxiang Ding}
  \email{dingcx@ahut.edu.cn}
 \affiliation{School of Science and Engineering of Mathematics and Physics, Anhui University of Technology, Maanshan 243002, China }
\author{Henk W. J. Bl\"ote}
 \affiliation{Instituut Lorentz, Leiden University, P.O. Box 9506, 2300 RA Leiden, The Netherlands}
 \author{Youjin Deng}
\email{yjdeng@ustc.edu.cn}
 \affiliation{Hefei National Laboratory for Physical Sciences at Microscale,
 Department of Modern Physics, University of Science and Technology of China, Hefei, 230027, China}
\date{\today}
\begin{abstract}
We study the $q$-state Potts model on the simple cubic lattice with ferromagnetic interactions
in one lattice direction, and antiferromagnetic interactions in the two other directions.
As the temperature $T$ decreases, the system undergoes a second-order phase transition 
that fits in the universality class of the 3D O($n$) model with $n=q-1$.
This conclusion is based on the estimated critical exponents, and histograms of the order parameter.
At even smaller $T$ we find, for $q=4$ and 5, a first-order transition to a phase with a different
type of long-range order.
This long-range order dissolves at $T=0$, and the system effectively reduces to a 
disordered two-dimensional Potts antiferromagnet.
These results are obtained by means of Monte Carlo simulations and finite-size scaling.
\end{abstract}
\maketitle 
\section{Introduction}
According to the hypothesis of universality, critical phase transitions fall in classes
determined by spatial dimensionality and symmetry of the order parameter. 
The latter is usually reflected by the degeneracy of the ground state of the Hamiltonian. 
However, for certain systems at criticality, a higher symmetry may emerge in the order parameter,
and the associated critical behavior may become very rich.

Examples of this phenomenon are known, in particular in two dimensions (2D).
In the $q$-state clock model~\cite{clock2D}, the spins are confined to a plane and take $q$ 
discrete directions $\phi=2 \pi n/q$. The ferromagnetic ground state is $q$-fold degenerate,
reflecting the $Z_q$ symmetry.
For $q \geq 5$, however, the 2D clock model exhibits a Berezinskii-Kosterlitz-Thouless (BKT)
transition as the temperature $T$ is lowered, and quasi-long-range order with continuous
O(2) symmetry emerges at $T<T_c$, just as in the rotationally invariant XY model.
Emergent symmetries are also found in many other physical systems, including 
a spin ice system~\cite{ice}, deconfined quantum critical points~\cite{quandec},
high-$T_{\rm c}$ superconductors~\cite{SCZhang} and so forth~\cite{spsym},
and are often accompanied by very interesting phenomena. For example, 
in the spin ice~\cite{ice}, emergent SU(2) symmetry leads to an unusual phase transition
with a jump in the order parameter, which is a feature of discontinuous transitions, whereas the 
the domain wall tension vanishes, which is a feature of continuous transitions.
In a class of models with $Z_2$ and U(1) symmetry~\cite{spsym}, emergent supersymmetry at the Ising-BKT 
multicritical point leads to new critical behavior, with unusual scaling of the correlation length.

The Potts model~\cite{Potts} has spins with $q$ values $\sigma=1,2,\cdots,q$ that interact as
$K \delta_{\sigma_i,\sigma_j}$, reflecting permutation symmetry.
The antiferromagnetic $q=3$ model on the simple cubic lattice breaks an effective $Z_6$ symmetry
at low temperatures, and O(2) symmetry emerges at the critical point~\cite{3sAFP,clock6s3D}.
In line with these findings, a BKT transitions with emergent O(2) symmetry
can arise on simple cubic lattices with a finite thickness~\cite{3slayer}.
For similar models with $q>3$ one might thus expect emergent O($n$) symmetry, {\em i.e.}, isotropy
in $n$ dimensions. While this scenario is consistent with numerical results~\cite{4sAFPMC},
the situation is not entirely clear~\cite{4sAFPRG,4sAFPMC}.

It has been hypothesized~\cite{SCZhang} that in high-$T_{\rm c}$ superconductors,
the magnetic and superconducting degrees
of freedom can merge into a critical state with effective SO(5) symmetry.  However, as argued by Fradkin
{\em et al}.~\cite{Fr}, the symmetry of the corresponding O(5) fixed point is easily broken since the
components of the order parameter are inequivalent on the microscale. This applies as well to other
systems~\cite{Fr} for which higher emergent symmetries had been proposed.

A different critical behavior occurs in two-dimensional systems with mixed interactions---i.e.,
ferromagnetic (FM) in one direction and antiferromagnetic (AF) in the other. 
The  $q=3$ mixed Potts model on the square lattice undergoes a BKT-like transition, and O(2) symmetry emerges
in the low-temperature range~\cite{mixedsquare}.  Rich phenomena also occur
in the the mixed Ising ($q=2$) model on the multi-layered triangular lattice~\cite{mixedIsing}.

In this paper, we study $q$-state mixed Potts models on the simple cubic lattice, 
with FM couplings in the $z$ direction and AF couplings in the $x-y$ plane.
Using cluster-type Monte Carlo algorithms, we find continuous phase transitions for
$2 \leq q \leq 6$ as the temperature $T$ is lowered. 
The critical behavior of these systems is consistent with O($n$) universality in 3D, with $n=q-1$.
This result may hold more generally, {\em i.e.}, for at least some $q >6$.
To our knowledge, such an emergent O($n$) symmetry in the 3D Potts model has,
apart from the $q=3$ antiferromagnet, not been reported in literature.  

\section{Model, Algorithm, critical points, and critical exponents}
The reduced Hamiltonian of the mixed Potts model is 
\begin{equation}
\label{e:Hamiltonian}
\mathcal{H}=K \sum\limits_z \sum\limits_{\langle i,j\rangle}\delta_{\sigma_{i,z},\sigma_{j,z}}-
                     K \sum\limits_z \sum\limits_i \delta_{\sigma_{i,z},\sigma_{i,z+1}} \; ,
\end{equation}
where the spins take $q$ values $\sigma=1,2,\cdots,q$.
With $K>0$, the sum in the first term, taken over all nearest-neighbor sites
$\langle i,j \rangle$ in layer $z$, defines AF couplings.
The second term defines FM couplings in the $z$ direction.
We refer to the temperature as $T=1/K$.
 
Cluster Monte Carlo methods are very effective for simulation of FM
lattice models~\cite{SW}, while the efficiency of the Wang-Swendsen-Koteck\'{y} (WSK)
algorithm~\cite{3sAFP} for AF Potts models depends on the lattice type and temperature.
For mixed interactions, we apply a single-cluster algorithm merging elements of
the Wolff method for FM models~\cite{Wolff}) and the WSK algorithm.
A combination with the geometric cluster algorithm~\cite{g1,g2,g3}, which employs
lattice symmetries, is still needed for effective simulations of $L^3$
systems up to $L=128$ at sufficiently low temperatures. In addition we applied Metropolis sweeps.

The sampled observables include the staggered susceptibility $\chi_{\rm s}$, the uniform susceptibility
$\chi_{\rm u}$, their Binder ratios $Q_{\rm u}$ and $Q_{\rm s}$, and the specific heat $C_{\rm v}$:
\begin{eqnarray}
\chi_{\rm s}&=&V\langle \mathcal{M}_{\rm s}^2\rangle, {~~~} Q_{\rm s}=\frac{\langle
\mathcal{M}_{\rm s}^4\rangle}{\langle \mathcal{M}_{\rm s}^2\rangle^2},\\
\chi_{\rm u}&=&V\langle \mathcal{M}_{\rm u}^2\rangle, {~~~} Q_{\rm u}=\frac{\langle
\mathcal{M}_{\rm u}^4\rangle}{\langle \mathcal{M}_{\rm u}^2\rangle^2},\\
C_{\rm v}&=&V(\langle \mathcal{E}^2\rangle-\langle \mathcal{E}\rangle^2)/T^2,
\end{eqnarray}
where $V=L^3$ is the system volume,
and $\mathcal{E}$ the energy density.
$\mathcal{M}_{\rm s}^2$ and $\mathcal{M}_{\rm u}^2$ are defined as 
\begin{eqnarray}
\mathcal{M}_{\rm s}^2&=&\frac{q}{q-1}\sum\limits_{p=1}^q(\rho_{a,p}-\rho_{b,p})^2 
\label{ms} \\
\mathcal{M}_{\rm u}^2&=&\frac{q}{q-1}\sum\limits_{p=1}^q(\rho_{a,p}+\rho_{b,p})^2-\frac{1}{q-1}
\label{mu}
\end{eqnarray}
with $\rho_{k,p}$ (with $k=a,b$) the density of state-$p$ spins on sublattice $k$, namely
\begin{eqnarray}
\rho_{k,p}&=&\frac{1}{V}\sum\limits_{\vec{r}\in k}\delta_{\sigma_{\vec{r}},p}\label{rho}
\end{eqnarray}
with spin coordinates $\vec{r}=(x,y,z)$. The sublattice $k$ is defined by the parity of $x+y$.
The $\mathcal{M}_s$ and $\mathcal{M}_u$ are the order parameters of the model, exposing a possible
symmetry breaking of the model. It should be noted that, in the AF Potts model, the type of order 
in the low-temperature phases depends essentially on entropy effects, apart from the energy effect.
For example, the low temperature phase of the 3-state AF Potts model on the simple cubic lattice
displays long-range order with one sublattice frozen in one of the Potts states, 
while the spins on the other sublattices are free to randomly take one of the other Potts states \cite{3sAFP}.
The maximal entropy of the latter sublattice explains the existence of this type of state. The
phase transition to this state is thus, at least in part, entropy-driven, similar to behavior
found for certain two-dimensional $q=4$ Potts antiferromagnets \cite{entropy-driven}. 

\begin{figure}[htpb]
\includegraphics[scale=0.75]{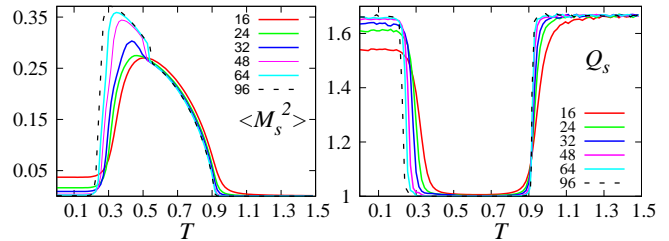}
\caption{(Color online) Squared staggered magnetization density  $\langle \mathcal{M}_{\rm s}^2 \rangle $ (left)
and dimensionless ratio $Q_{\rm s}$ (right) versus $T$ for the 4-state mixed Potts model.
The staggered magnetization becomes non-zero below $T_{\rm c} \approx 0.91$.   
Near $T\approx 0.55$, $\langle \mathcal{M}_{\rm s}^2 \rangle $
displays a jump, signaling a first-order phase transition.
For $T < 0.2$, the correlation length in the $z$ direction exceeds the system size, 
and crossover occurs to the disordered $T=0$ ground states.}
\label{chisQs4}
\end{figure}

\begin{figure}[htpb]
\includegraphics[scale=0.75]{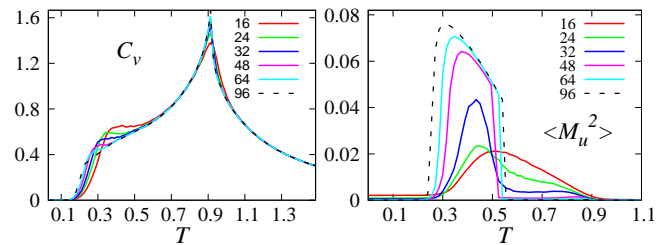}
\caption{(Color online) Specific heat $C_{\rm v}$ (left) and squared magnetization 
density  $\langle \mathcal{M}_{\rm u}^2 \rangle $ (right) versus $T$ for the 4-state mixed Potts model. 
The specific heat $C_{\rm v}$ displays a cusp at $T_{\rm c} \approx 0.91$, while
$\langle \mathcal{M}_{\rm u}^2 \rangle $ develops a discontinuity at the first-order transition
$T_{\rm c}' \approx 0.55$. }
\label{Cvchiu4}
\end{figure}

We have investigated the model~(\ref{e:Hamiltonian}) for $q=3,4,5$ and 6, with
periodic boundary conditions. The procedure involved three steps, specified here for $q=4$.
\begin{ruledtabular}
\begin{table*}[htbp]
\caption{Critical points and critical exponents of the $q$-state mixed Potts model on the simple cubic
lattice; $a$ reflects a conservative average of some recent results given in Refs.~\onlinecite{Ising3D} 
and \onlinecite{TIM}.
The critical exponents $y_t$ and $y_h$ of the recent results for the 3D O($n$) model, which are taken
from Ref.~\onlinecite{loop3D}, are also listed here for comparison. }
\begin{tabular}{r l l l l | l l l}
 \multicolumn{5}{c|}{$q$-state mixed Potts model}&\multicolumn{3}{c}{O($n$) model}\\
$q$&$T_{\rm c}$&$y_t$&$y_h$& $y_u$ & $n$ &$y_t$ &$y_h$ \\
\hline
$2$&2.2557616(8)$^a$&1.5873(1)$^a$&2.481846(15)$^a$&$=y_t$&1&1.588(2)&2.483(3)\\
$3$&1.36086(1)  &1.488(4)&2.483(2)&1.754(4) &2&1.488(3)&2.483(2)\\
$4$&0.91381(1)  &1.402(6)&2.485(3)&1.787(5) &3&1.398(2)&2.482(2)\\
$5$&0.64116(2)  &1.33(2) &2.482(3)&1.806(5) &4&1.332(7)&2.483(2)\\
$6$&0.44545(2)  &1.30(3) &2.485(3)&1.837(4) &5&1.275(12)&2.483(3)\\
\end{tabular}
\label{pottstab}
\end{table*}
\end{ruledtabular}
First, we simulate for several $L$ at a number of temperatures $T$ taken in a wide range.
Each data point is based on $2 \times 10^6$ Monte Carlo steps (MCSs). Each MCS consists of 5 Wolff-cluster
updates of the WSK type, 5 geometric-cluster updates, 5 Metropolis sweeps, and data sampling. 
Each different simulation uses a different random seed, and starts from a random initial configuration,
after which about $5\times 10^5$ data samples are discarded to allow for equilibration of the system.
These simulations are distributed over different CPU cores.
After their completion, the resulting data are collected, and the averages of the physical variables
and their error bars are calculated. 
Plots of $\langle \mathcal{M}_{\rm s}^2\rangle$ and $Q_{\rm s}$ in Fig.~\ref{chisQs4} yield an 
approximate critical point $T_{\rm c}\approx 0.91$.
This is also seen in the scaling of the specific heat, in the left panel of Fig.~\ref{Cvchiu4}.

Next, in order to determine the critical exponents of this phase transition
we simulate near the estimated $T_{\rm c}$, with  $1.6\times10^7$ MCS
taken at each data point.
The $Q_{\rm s}$ data scale as 
\begin{eqnarray}
Q_{\rm s}&=&Q_{\rm s0}+\sum\limits_{k=1}a_k(T-T_{\rm c})^{k}L^{ky_t}+bL^{y_1},
\end{eqnarray}
where $y_t>0$ is the thermal exponent, $y_1<0$ is the
correction-to-scaling exponent, and $Q_{\rm s0}$, $a_k$ and $b$ are unknowns.
A least-squares fit of this formula to the data yields $T_{\rm c}=0.91381(1)$,
and $y_t=1.402(6)$.

In the last step, we simulate at $T_{\rm c}$ and fit 
the data of $\chi_{\rm s}$ by
\begin{eqnarray}
\chi_{\rm s}=L^{2y_h-d}(a+bL^{y_1}), \label{chisfss}
\end{eqnarray}
with spatial dimensionality $d=3$. The magnetic exponent follows as $y_h=2.485(3)$. 
The uniform susceptibility $\chi_{\rm u}$ was also
fitted by  Eq.~(\ref{chisfss}), with $y_h$ replaced by another magnetic exponent $y_{\rm u}$.
This fit gives $y_{\rm u}=1.787(5)$.

The results for $q=3$, $5$ and 6 used the same procedure. The critical points and exponents
are listed in Table \ref{pottstab}. A comparison with the exponents $y_t$ and $y_h$ of the 
O($n$) model~\cite{loop3D} shows that the phase transition of the $q$-state mixed
Potts model fits the universality class of the O($n$) model with $n=q-1$.

\section{Histogram of the O(n) symmetry}
The remarkable emergence of 3D O($n=q-1$) universality in these models invites the
construction of order parameter histograms, by representing the spins as vectors.
These $q$ vectors are symmetrically distributed in $(q-1)$-dimensional space, 
such that their scalar product matches the pair potential of Eq.~(\ref{e:Hamiltonian}), which 
can be written as 
\begin{eqnarray}
\vec{e}_\sigma \cdot \vec{e}_\sigma&=&1, ~~~({\rm no~summation on}~\sigma)\\
\vec{e}_\sigma \cdot \vec{e}_{\sigma^\prime}&=&-\frac{1}{q-1} ~~~ (\sigma\ne\sigma^\prime)
\end{eqnarray}
For example, in the case of $q=3$ these vectors span a regular triangle,
{\em i.e.}, $\vec{e}_\sigma=(\cos\theta,\sin\theta)$ with $\theta=2\sigma\pi/3$.
For $q=4$, the vectors are three-dimensional ones, and span a regular tetrahedron: 
\begin{eqnarray}
\vec{e}_\sigma&=&(+1,+1,+1)/\sqrt{3}, \mbox{~for~} \sigma=1;\nonumber\\
            &=&(+1,-1,-1)/\sqrt{3}, \mbox{~for~} \sigma=2;\nonumber\\
            &=&(-1,-1,+1)/\sqrt{3}, \mbox{~for~} \sigma=3;\nonumber\\
            &=&(-1,+1,-1)/\sqrt{3}, \mbox{~for~} \sigma=4.\nonumber \, .
\end{eqnarray}
Based on this vector representation, the magnetization is sampled separately for sublattices $a$ and $b$.
This yields the components
of $\mathcal{M}_{\rm s}$ and $\mathcal{M}_{\rm u}$ in Eqs.~(\ref{ms}) and (\ref{mu})
\begin{equation}
\mathcal{M}_{\rm s}=\mathcal{M}_a-\mathcal{M}_b\,,~~~
\mathcal{M}_{\rm u}=\mathcal{M}_a+\mathcal{M}_b\,.
\end{equation}

\begin{figure}[htpb]
\includegraphics[scale=0.48]{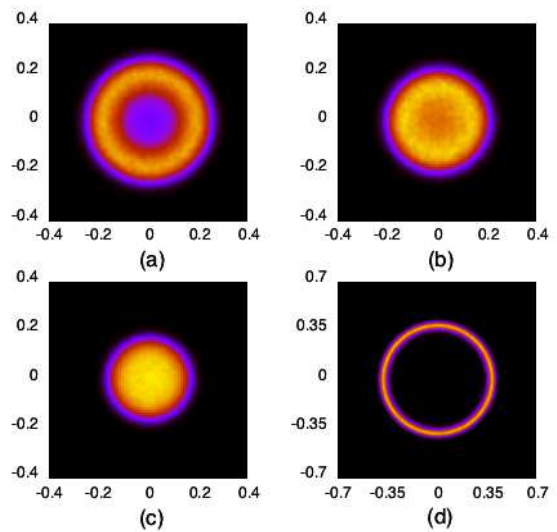}
\caption{(Color online)
Histograms of $\mathcal{M}_{\rm s}$ for (a) $q=3$ at criticality;
(b) $q=4$ at criticality; (c) $q=5$ at criticality; and (d)
$q=3$ at $T=1.30<T_{\rm c}=1.36$.}
\label{histo}
\end{figure}

Figures~\ref{histo}(a)-(c) display the histograms of the staggered magnetization for $L=32$ systems,
projected on two Cartesian axes, for $q=3$, 4, and 5 respectively. The histograms are the same for
any choice of the axes.
The apparent isotropy shows the emergent O($q-1$) symmetry. For $q=3$, the symmetry persists
in a finite system for a range below $T_c$, as shown in Fig.~\ref{histo}(d).
\begin{figure}[htpb]
\includegraphics[scale=0.65]{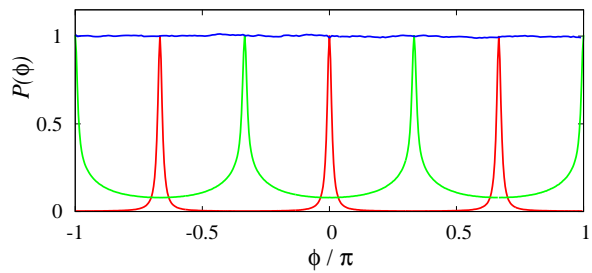} \vspace*{-8mm}
\caption{(Color online) Histogram of the orientation $\phi$ of $\mathcal{M}_{\rm s}$
of a critical $L=32$, $q=5$ mixed Potts model (blue line). It is,
modulo $\pi$, given by $\phi\equiv \arctan(\mathcal{M}_{\rm sy}/ \mathcal{M}_{\rm sx})$.
It is compared with analogous data for $\mathcal{M}_{\rm u}$
of $q=3$ Potts ferromagnets at $T_c$ with $d=2$ ($L=240$, red peaks) 
and with $d=3$ ($L=32$, green peaks). The $d=3$ transition is weakly
discontinuous. Its data, shifted by $\pi/3$ for clarity, display a 
background from the coexisting disordered phase.
The histograms are rescaled to match each other.}
\label{histo53}
\end{figure}
Figure \ref{histo53} shows that Potts ferromagnets behave differently.
It compares the histogram of the orientation $\phi$ of the staggered magnetization 
of the $q=5$ mixed Potts model, projected on the $x,y$ plane, to
similar plots for the magnetization of $q=3$ ferromagnets with $d=2$ and 3.

\section{First order transition of the model}
\begin{figure}[htpb]
\includegraphics[scale=0.38]{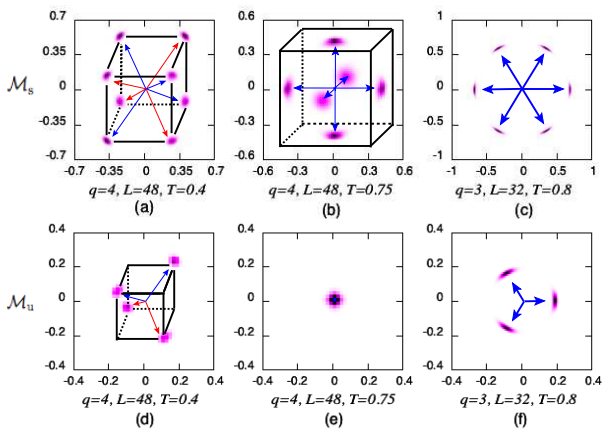}
\caption{(Color online) Histogram of $\mathcal{M}_{\rm s}$ and $\mathcal{M}_{\rm u}$ for the 4-state
and 3-state mixed Potts models. 
Lines are added to guide the eyes, especially to a 3D impression. The arrows describe the symmetry of 
$\mathcal{M}_{\rm s}$ and of $\mathcal{M}_{\rm u}$.}
\label{symmetry}
\end{figure}

\begin{figure}[htpb]
\includegraphics[scale=0.75]{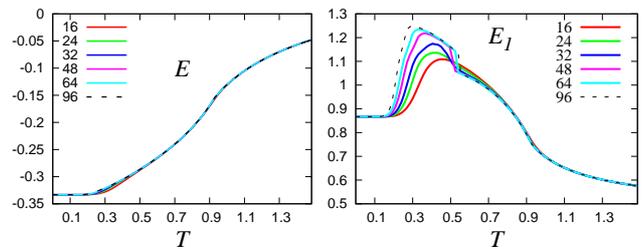}
\caption{(Color online) Left: energy density $E$ of the 4-state mixed Potts model. Right: energy-like quantity density
$E_1$, defined in the text, of the 4-state Potts model.
The behavior of the curves of $E_1$ with increasing system sizes clearly indicates the existence of a discontinuity at
$T_c^\prime\approx 0.55$ in the thermodynamic limit. }
\label{En}
\end{figure}

In the range below $T_c$, the 4-state mixed Potts model displays jumps in $\mathcal{M}_{\rm s}$ and
$\mathcal{M}_{\rm u}$, near $T_{\rm c}^\prime\approx 0.55$ in Figs.~\ref{chisQs4} and \ref{Cvchiu4}.
In the middle range $T_{\rm c}^\prime<T<T_{\rm c}$, $\langle \mathcal{M}_{\rm u}^2\rangle$
vanishes for $ L \rightarrow \infty$,
while $\langle \mathcal{M}_{\rm s}^2 \rangle$ converges to a nonzero value. 
In contrast, both $\langle \mathcal{M}_{\rm s}^2\rangle$ and $\langle \mathcal{M}_{\rm u}^2\rangle$
converge to nonzero values in the low $T$ range $0<T<T_{\rm c}^\prime$. 
Thus different symmetries are broken on the two sides of $T_{\rm c}^\prime$.
The histograms of $\mathcal{M}_{\rm s}$ and $\mathcal{M}_{\rm u}$ show
a broken $Z_2\times S_4$ symmetry for $0<T<T_{\rm c}$, with
$Z_2$ for the permutation of the two sublattices, and $S_4$ for the symmetric group for
the four-state Potts model.
In a typical configuration at $0<T<T_{\rm c}'$, one Potts state dominates one sublattice, and the remaining states
randomly occur on the other sublattice.
This implies the breaking of the corner-cubic symmetry described by the four
vectors $\vec{e}_\sigma$ for $q=4$, preceded by a sublattice sign $\pm$.
In a typical configuration at $T_{\rm c}^\prime<T<T_{\rm c}$, one sublattice is dominated by
two random spin states, and the other sublattice by the other states.
The staggered magnetization vector then points at one face of a cube,
signaling a broken face-cubic symmetry. 
The histograms of $\mathcal{M}_{\rm s}$ are shown in Fig.~\ref{symmetry}(a) for $0<T<T_{\rm c}$ and
Fig.~\ref{symmetry}(b) for $T_{\rm c}^\prime<T<T_{\rm c}$, clearly displaying the broken corner-cubic
and face-cubic symmetries.
Similarly, Figs.~\ref{symmetry}(d) and \ref{symmetry}(e) show the histogram of $\mathcal{M}_{\rm u}$
for $0<T<T_{\rm c}$,
and $T_{\rm c}^\prime<T<T_{\rm c}$, respectively. The transition at $T_{\rm c}^\prime$ is not well
visible in the ordinary energy density $E=\langle \mathcal{E}\rangle$, but it is exposed by the
energy-like quantity $E_1=\langle \mathcal{E}_1 \rangle$ based on the next-nearest neighbor correlations
in the $x$-$y$ planes, expressed as
\begin{eqnarray}
  \mathcal{E}_1=\frac{1}{L^3}\sum\limits_{<<i,j>>} \delta_{\sigma_i,\sigma_j},
\end{eqnarray}
where $<<i,j>>$ denote the next-nearest neighboring sites in the $X-Y$ planes. 
Figure \ref{En} shows the curves of $E$ versus $T$ and the curves of $E_1$ versus $T$, 
the curves of $E$ are featureless for this transition,
but the curves of $E_1$ at $T_c^\prime$ for large system sizes show an obvious energy gap.
This result reflects the stronger next-nearest neighbor correlations in the corner-cubic phase.

The 3-state mixed Potts model breaks the $Z_2\times S_3$ symmetry in the whole range $0<T<T_{\rm c}$,
as shown in Figs.~\ref{symmetry}(c) and (f).
But for the $q=5$, another similar discontinuous transition appears at $T_{\rm c}^\prime\approx0.4$, 
which  breaks a $Z_2\times S_5$ symmetry in the low-$T$ range $0<T<T_{\rm c}^\prime$.
The degeneracy in the intermediate range $T_{\rm c}^\prime<T<T_{\rm c}$ is $2\times C^5_2 = 20$,
where $C^5_2$ denotes the binomial coefficient.
In a typical configuration at $T_{\rm c}^\prime<T<T_{\rm c}$, the spins on one sublattice randomly
take two states, and the other spins randomly take the remaining three states. 

A similar discontinuous transition may occur in the 6-state mixed Potts model. 
We observed that the $Z_2 \times S_6 $ symmetry is broken at low $T$, 
and that another ordered phase exists at intermediate $T$. But we did not find a jump 
in $\mathcal{M}_{\rm s}$ or $\mathcal{M}_{\rm u}$ for systems up to $L=96$.
This may still be due to a strong finite-size effect. 

\begin{figure}[htpb]
 \includegraphics[scale=0.4]{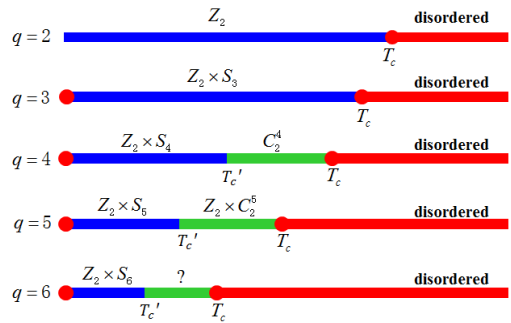}
\caption{(Color online) Phase diagram of the $q$-state mixed Potts model. The multiplicities
of the intermediate phases are expressed with help of the binomial coefficients $C^4_2$ and $C^5_2$.
The question mark means that the phase has not been unambiguously identified.
The models are disordered at $T=0$ for $q > 3$.}
\label{pt}
\end{figure}

At zero temperature the model reduces to a $T=0$ square-lattice AF Potts model~\cite{SokalSquare},
which is N\'{e}el ordered for $q=2$, critical for $q=3$, and disordered for $q>3$.
Figure~\ref{pt} summarizes the phase behavior of the $q$-state mixed Potts models.

\section{Conclusion and discussion}
In summary, our results indicate that $q$-state Potts models on the simple cubic lattice with mixed FM
and AF interactions display continuous phase transitions, with critical
exponents in the O($n$) universality class with $n=q-1$. 
The order parameter displays this emergent symmetry at criticality.
In the low temperature ranges of the $q=4$ and 5 models, perhaps
also for $q>5$, a discontinuous transition occurs between two ordered phases. 
For $T \to 0$, the model crosses over to the $d=2$ square-lattice 
AF Potts model, which is disordered for $q >3$. 

Although the temperature is the only variable, the $q$-state mixed 
Potts model displays diverse and enigmatic phenomena.
The O($n$) symmetry is not at all obvious in the Hamiltonian, but it nevertheless emerges, and controls the 
critical properties of the continuous phase transition at $T_c$. 
In the sense of universality, the 3-state mixed Potts model is similar to the O(2) model with a $Z_6$
perturbation\cite{sandvik2007}, which also displays an emergent O(2) symmetry at criticality.
For $q>3$, the mixed Potts model  has a low-temperature ordered phase that spontaneously
breaks the $(q-1)$-dimensional  face-cubic symmetry.
The histogram of the order parameter at criticality shows an emergent O($n=q-1$) symmetry,
and the estimated thermal exponent $y_t$ is  a decreasing function of $q$, consistent with the O($q-$1) universality.
In the analogous case of the pure antiferromagnetic $q=4$ Potts model, the low-temperature ordered phase 
also breaks the 3D face-cubic symmetry and that the Monte Carlo simulation up to $L=96$ also
yields critical exponents consistent with the O(3) universality class~\cite{4sAFPMC}.
Since the cubic perturbation is expected to be relevant for the O(3) model~\cite{CPV},
one cannot fully exclude that for $q=4$, these phase transitions are of weak first order or belong to another universality class.
However, in either case the question still remains why the effects of this perturbation are invisible
in our analysis of finite systems. 
On the basis of our systematic study of the $q$-state mixed Potts models, we conclude that 
any symmetry-lowering perturbations of the emergent symmetries are strongly suppressed, 
allowing the possibility that the ordering transitions fit exactly in the O($q-1$)
universality classes.

Finally, we mention that the mixed Potts model resembles a square-lattice quantum Potts antiferromagnet in a transverse
field~\cite{qp1,qp2,qp3}. The $z$ dimension in the classical model  corresponds
with imaginary time in the Suzuki-Trotter formulation of the quantum model.
The present series of mixed Potts models may provide a simple example where quantum fluctuations
give rise to rich behavior.

\section{Acknowledgment}
We thank Cristian D. Batista for valuable discussions. 
This work is supported by the National Science Foundation of China
 (NSFC) under Grant Nos. 11205005  and 11275185, 
and by the Anhui Provincial Natural Science Foundation under Grant No. 1508085QA05.

\end{document}